\documentclass{elsart_TL}
\usepackage{graphicx}
\usepackage{amssymb}

\begin{document}

\begin{frontmatter}

%\title{Uniaxial pressure dependencies of the phase transition temperatures in
%multiferroic GdMnO$_3$}
\title{Uniaxial pressure dependencies of the phase transitions in GdMnO$_3$}

\author[Colonia]{J. Baier}
\author[Colonia]{D. Meier}
\author[Colonia]{K. Berggold}
\author[Aug]{J. Hemberger}
\author[Mos]{A. Balbashov}
\author[Colonia]{J.A. Mydosh}
\author[Colonia]{T. Lorenz\corauthref{Lorenz}}
\ead{lorenz@ph2.uni-koeln.de}

\address[Colonia]{II.\,Physikalisches Institut, Universit\"{a}t zu
K\"{o}ln, Z\"{u}lpicher Str. 77, 50937 K\"{o}ln, Germany}
\address[Aug]{Experimentalphysik V, Institut f\"{u}r Physik,
 University of Augsburg, 86135 Augsburg, Germany}
\address[Mos]{Moscow Power Engineering
 Institute, 105835 Moscow, Russia}

\corauth[Lorenz]{Corresponding author:}

\begin{abstract}

GdMnO$_3$ shows an incommensurate antiferromagnetic order below
$\simeq 42$~K, transforms into a canted A-type antiferromagnet
below $\simeq 20$~K, and for finite magnetic fields along the $b$
axis ferroelectric order occurs below $\simeq 12$~K. From
high-resolution thermal expansion measurements along all three
principal axes, we determine the uniaxial pressure dependencies
of the various transition temperatures and discuss their
correlation to changes of the magnetic exchange couplings in
$R$MnO$_3$ ($R = {\rm La, \ldots , Dy}$).
\end{abstract}

%%%%%%%%%use  the \KEY command at the begin of keyword text%%%%%%%%%
%Please supply one or more relevant PACS-1996 classification codes
%(http://publish.aps.org/PACS/96pacs.htmland) and about 5 keywords
%of your own choice for indexing purposes.
%You can see a list of already used keywords for JMMM at
%http://authors.elsevier.com/JournalDetail.html?PubID=505704&Precis=KIND

\end{frontmatter}

\begin{figure}[b]
\begin{center}
\includegraphics[clip,width=.48\textwidth]{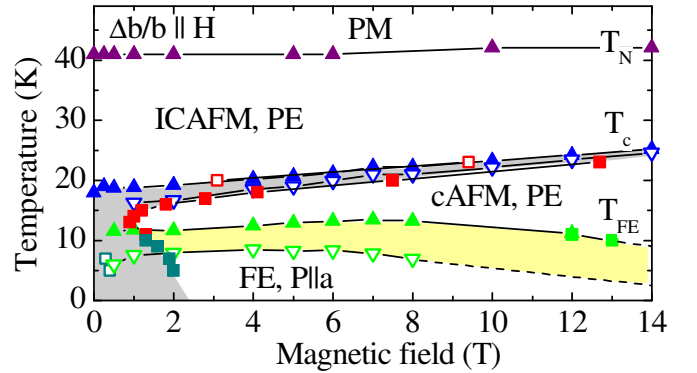}
\end{center}
\caption{Phase diagram of GdMnO$_3$ based on thermal expansion
($\blacktriangle \,\, \triangledown $) and magnetostriction
measurements ({\small $\blacksquare\,\, \square$}) for magnetic
fields applied along the $b$ axis. Filled and open symbols are
obtained with increasing and decreasing $T$ or $H$, respectively.
Shaded areas signal regions of strong hysteresis (see also
Ref.~\cite{baier06a}).} \label{fig:Phadi}
\end{figure}

GdMnO$_3$ is one of the multiferroic rare-earth manganites
\textsl{R}MnO$_3$ with \textsl{R} = Gd, Tb, and Dy where magnetic
and ferroelectric order coexist.\cite{goto04a} While for
\textsl{R} = Tb and Dy the multiferroic behavior is already
present in zero magnetic field,\cite{kimura05a} a finite magnetic
field along the $b$ axis is necessary to induce multiferroicity
in GdMnO$_3$. Recently, we have determined the $H$-$T$ phase
diagram for different field orientations by measurements of the
uniaxial thermal expansion and magnetostriction along the $a$,
$b$, and $c$ axis of GdMnO$_3$.\cite{baier06a} As shown in
Fig.~\ref{fig:Phadi}, there is a transition from the paramagnetic
(PM) to an incommensurate antiferromagnetic phase (ICAFM) around
$T_{\rm N}\simeq 42$~K. For fields above $\simeq 2$~T, the ICAFM
transforms into a canted A-type antiferromagnet (cAFM) around
$T_{\rm c} \simeq 20$~K, and a transition to a ferroelectric (FE)
phase occurs around $T_{\rm FE}\simeq 12$~K. The notations
'ICAFM' and 'cAFM' should be treated with some caution, because
the magnetic structure of GdMnO$_3$ has not yet been determined
unambiguously.

With increasing field-strength, $T_{\rm N}$ shows a weak, and
$T_{\rm c}$ a moderate, increase while $T_{\rm FE}$ passes
through a broad maximum around 7~T. The PM-to-ICAFM transition is
of second order, while the other two transitions are of
first-order with very strong hysteresis, which is most pronounced
in the low-field region.\cite{baier06a} The thermal expansion
measurements do not only allow to detect the transition
temperatures $T_{\rm c}$. Together with specific heat data, it is
also possible to determine their dependence on uniaxial pressure
$p_i$ via
\begin{equation}
\frac{\partial T_{\rm c}}{\partial p_i} = V_m \frac{\Delta
L_i/L_i}{\Delta S}
 \hspace*{5mm} \mbox{ or } \hspace*{5mm}
 \frac{\partial T_{\rm c}}{\partial
p_i} = V_m T_{\rm c} \frac{\Delta \alpha_i}{\Delta c}
 \,\,\, .
 \label{Ehr}
\end{equation}
The first expression is the Clausius-Clapeyron and the second one
is the Ehrenfest relation, which are valid for first- and
second-order phase transitions, respectively. $V_m$ denotes the
molar volume, while $\Delta S $ and $\Delta L_i$ are the
respective discontinuous changes of the entropy $S$ and the
sample length $L_i$. For a second-order transition $S$ and $L_i$
change continuously, but the specific heat $c=T\partial
S/\partial T$ and the thermal expansion coefficients
$\alpha_i=1/L_i \cdot
\partial L_i/\partial T$ exhibit jumps at $T_{\rm c}$.

\begin{figure}[t]
\begin{center}
\includegraphics[clip,width=.48\textwidth]{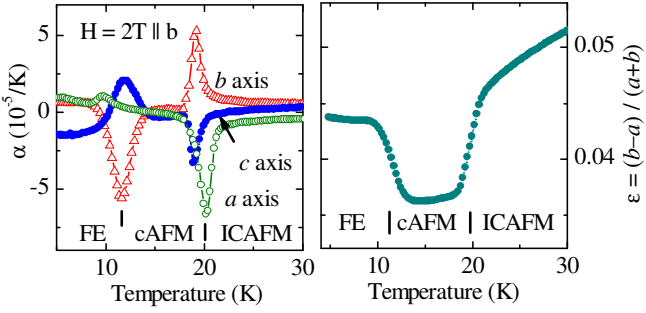}
\end{center}
\caption{Left: Thermal expansion of GdMnO$_3$ along the $a$, $b$,
and $c$ axis measured in a magnetic field of 2~T parallel to $b$.
Right: Orthorhombic distortion calculated via integration of
$\alpha_a$ and $\alpha_b$ (see text).} \label{fig:alp}
\end{figure}

Fig.~\ref{fig:alp} displays $\alpha_i$ of the $a$, $b$, and $c$
axes. In all three cases a magnetic field of 2~T has been applied
along the $b$ axis. We have chosen this configuration, because
only for fields above $\simeq 2$~T along $b$ can all three
transitions be observed in GdMnO$_3$. It would be even better to
use a larger field. However, the canted moment of the cAFM phase
is parallel to $c$, and the torque $\overrightarrow{M} \times
\overrightarrow{H}$ tends to rotate the sample in fields applied
along $a$ or $b$. Along all three axes $\alpha_i$ shows very large
anomalies at the ICAFM-to-cAFM and at the cAFM-to-FE transition.
The anomalies for the different $i$ occur at slightly different
temperatures. We suspect that these differences arise from a
misorientation of the sample with respect to the field direction.
In the transverse configurations, part of the misorientation is
due to the above-mentioned torque, since the sample is not
completely fixed during the measurement. This problem would
hamper a quantitative determination of the pressure dependencies.
In the following we will, however, restrict ourselves to a
qualitative discussion, since up to now specific-heat data are
only available for magnetic fields applied along
$c$.\cite{hemberger04b}

\begin{figure}[t]
\begin{center}
\includegraphics[clip,width=.48\textwidth]{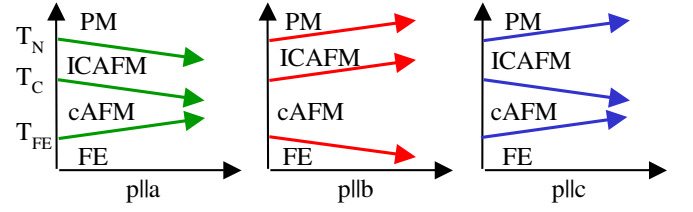}
\end{center}
\caption{Dependencies of the transition temperatures of GdMnO$_3$
on uniaxial pressure along the $a$, $b$, and $c$ axis (see text).
} \label{fig:pres}
\end{figure}

Because any ordering decreases $S$, the anomalies $\Delta S$ and
$\Delta c$ have positive signs. Thus, the signs of $\partial
T_{\rm c}/\partial p_i$ are determined by the signs of $\Delta
L_i$ or $\Delta \alpha_i$.  A schematic overview of the expected
changes of $T_{\rm N}$, $T_{\rm c}$ and $T_{\rm FE}$ under
uniaxial pressure along $a$, $b$, and $c$ is given in
Fig.~\ref{fig:pres}. For the pressure-dependencies of $T_{\rm N}$
we used our previous data.\cite{baier06a} These can be compared
to the increasing structural distortion of the GdFeO$_3$ type as
a function of decreasing ionic radius of $R = {\rm La, \ldots
Dy}$ in the series \textsl{R}MnO$_3$.\cite{goto04a,kimura03b}
Starting from a cubic perovskite, the GdFeO$_3$ distortion can be
described by a rotation of the MnO$_6$ octahedra around the $c$
axis and a tilt around the $b$ axis. As a consequence, the $c$
axis continuously decreases and the orthorhombic distortion
$\varepsilon=(b-a)/(b+a)$ increases with decreasing radius of $R$.
From $R = {\rm La, \ldots, Eu}$, $T_{\rm N}$ of the cAFM ordering
decreases continuously, while $T_{\rm N}$ of the ICAFM order
slightly increases from Gd to Tb and decreases again from Tb to
Dy. In a simplified model~\cite{kimura03b}, the decreasing
$T_{\rm N}$ until Eu arises from a continuous decrease of the
ferromagnetic nearest-neighbor exchange $J_{\rm NN}^{\rm FM}$ in
the $ab$ plane, because of the decreasing Mn-O-Mn bond angle. In
addition, the antiferromagnetic next-nearest-neighbor exchange
$J_{\rm NNN}^{\rm AFM}$ becomes anisotropic. The competition of
$J_{\rm NN}^{\rm FM}$ and $J_{\rm NNN}^{\rm AFM}$ causes the ICAFM
order for $R={\rm Gd, \ldots, Dy}$, and certain types of helical
magnetic order may induce also a FE ordering.\cite{mostovoy05a}
At each transition of GdMnO$_3$, $p_a$ causes the opposite effect
as $p_b$. This anti-correlation just reflects the behavior of
$\varepsilon $. Using the room-temperature lattice constants, we
have also calculated $\varepsilon(T)$ by integration of
$\alpha_a$ and $\alpha_b$. As shown in Fig.~\ref{fig:alp}, the
ICAFM-to-cAFM (cAFM-to-FE) transition causes an abrupt decrease
(increase) of $\varepsilon$.

Within the above-mentioned model we interpret the uniaxial
pressure dependencies as follows: Pressure along $a$ increases
$\varepsilon$ and decreases $J_{\rm NN}^{\rm FM}$. As a
consequence, $T_{\rm N}$ and $T_{\rm c}$ decrease for $p_a$. As
discussed already in Ref.~\cite{baier06a}, the origin of the FE
order is not yet clear, but one may speculate that it arises from
some kind of helical order, which is induced by a magnetic field
along $b$. In this case, it is natural that $T_{\rm FE}$
increases for $p_a$, because $p_a$ destabilizes the cAFM order
and acts in the same direction as the (hypothetical)
magnetic-field influence. For pressure along $b$, the same
argumentation with inverted signs holds, because $\varepsilon$
decreases for $p_b$. Since pressure along $c$ will mainly shorten
the $c$ axis, it also acts in the same direction as a decreasing
radius of $R$. Therefore, one may expect that $p_c$ has a similar
influence as $p_a$. In fact, this is the case for the pressure
dependencies of $T_{\rm c}$ and $T_{\rm FE}$, but the increase of
$T_{\rm N}$ for $p_c$ cannot be explained by the above arguments.
Up to now we only considered the couplings within the $ab$ plane.
In order to establish any kind of magnetic ordering from the PM
phase, it is, however, necessary to have a finite coupling
$J_\perp$ along $c$. Thus, we attribute the increase of $T_{\rm
N}$ for $p_c$ to an increasing $J_\perp$, which overcompensates
the influence of the decreasing $J_{\rm NN}^{\rm FM}$ due to the
increasing distortion.

In summary, we measured the thermal expansion along all three
lattice directions of GdMnO$_3$. From these data we obtained how
the transition temperatures of the various phase transitions
change under uniaxial pressure, and how the uniaxial pressure
dependencies yield information about the relevant exchange
coupling of $R$MnO$_3$.

This work was supported by the DFG via SFB 608.

%We acknowledge fruitful discussions with ????

%This work was supported by the Deutsche Forschungsgemeinschaft
%via Sonderforschungsbereich 608.

%\vspace*{-5mm}

%\bibliographystyle{p:/bst/efk_english}
%\bibliography{p:/preload,p:/RMnO3}

\begin{thebibliography}{00}

\bibitem{goto04a}
T.~Goto {\it et al.},
\newblock Phys.\ Rev.\ Lett. {\bf 92}, 257201 (2004).

\bibitem{kimura05a}
T.~Kimura {\it et al.},
\newblock Phys.\ Rev.\ B {\bf 71}, 224425 (2005).

\bibitem{baier06a}
J.~Baier {\it et al.},
\newblock Phys.\ Rev.\ B {\bf 73}, 100402(R) (2006).

\bibitem{hemberger04b}
J.~Hemberger {\it et al.},
\newblock Phys.\ Rev.\ B {\bf 70}, 024414 (2004).

\bibitem{kimura03b}
T.~Kimura {\it et al.},
\newblock Phys.\ Rev.\ B {\bf 68}, 060403(R) (2003).

\bibitem{mostovoy05a}
M.~Mostovoy,
\newblock Phys.\ Rev.\ Lett. {\bf 96}, 067601 (2006).

\end{thebibliography}

\end{document}